%% file: wcprl.tex
\RequirePackage{lineno}
\documentclass[aps,prd,twocolumn,superscriptaddress,showpacs,preprintnumbers,amsmath,amssymb]{revtex4}

\usepackage{fancyhdr}
\usepackage{lineno}
\usepackage{slashed}

\usepackage{graphicx}
\usepackage{dcolumn}
\usepackage{bm}
\usepackage{longtable}

\begin{document}

\title{\boldmath First Measurement of the Production of a $W$ Boson in Association with a Single Charm Quark in $p\bar{p}$ Collisions at $\sqrt{s}=1.96~{\rm TeV}$}


\date{\today}
\begin{abstract}  
We present the first measurement of the production cross section of a $W$ boson with a single charm quark ($c$) in $p\bar{p}$ collisions at $\sqrt{s}=1.96~{\rm TeV}$, using soft muon tagging of $c$ jets. In a data sample of $\sim1.8$ fb$^{-1}$, recorded with the CDF II detector at the Fermilab Tevatron, we select events with $W+1$ or 2 jets. We use the charge correlation between the $W$ and the muon from the semileptonic decay of a charm hadron to extract the $Wc$ signal. We measure $\sigma_{Wc}\left(p_{Tc}>20~{\rm GeV}/c,~|\eta_c|<1.5\right)\times {\rm BR}(W\rightarrow \ell\nu) = 9.8\pm3.2~{\rm pb}$, in agreement with theoretical expectations.   
\end{abstract}
\input{July_2007_Authors_Visitors}
\maketitle

In the standard model (SM), the associated production of $W$ bosons with single charm quarks ($Wc$) in proton-antiproton collisions is described at leading order (LO) by the scattering of a gluon with a down, strange or bottom quark. Despite the dominant $d$-quark parton density function (PDF) the first process contributes only $\sim10$\% at the Tevatron, suppressed by the small Cabibbo-Kobayashi-Maskawa (CKM) matrix element $V_{cd}$. The contribution from $gb\rightarrow Wc$ is also heavily suppressed by $V_{cb}$ and the $b$ quark PDF. Therefore, about 90\% of the $Wc$ signal is produced by strange quark-gluon fusion and the production cross section is directly sensitive to the gluon and $s$-quark PDFs in the proton \cite{mangano}, at a scale of order of the $W$ mass ($M_W$), and to the element of the CKM matrix $V_{cs}$. Calculations of $W+{\rm heavy~quark}$ production are available at LO and next-to-leading order (NLO) in QCD \cite{keller}. From direct measurements, $|V_{cs}|$ is known with a precision of $\sim35$\% from $W^+\rightarrow c\bar{s}$, and of $\sim10$\% from $D$, $D_s$ decays \cite{pdg}, largely due to theoretical uncertainties. The $s$-quark PDF is known from neutrino-nucleon deep inelastic scattering experiments at momentum transfers, $Q^2$, about three orders of magnitude lower than $M_W^2$ \cite{nutev}. Moreover, a lack of constraints on the shape of the $s$-quark PDF has been shown to affect the calculation of the $Wc$ cross section by up to 20\%  \cite{lai}. The dependence on the factorization and renormalization scales in the NLO calculation yields an additional uncertainty of $\sim20$\% \cite{keller}. Overall, the uncertainty on the theoretical expectation of $Wc$ production is $\sim30$\%. Outside the SM, an exotic particle with high semileptonic branching ratio (BR) produced with a $W$ boson, as in the hypothesis investigated in \cite{giromini,superjets,d0sj}, could lead to a measured cross section of $Wc$ higher than expected. In addition, charged Higgs production via $c+\overline{s}\rightarrow H^+$ is sensitive the the strange PDF \cite{lai}. Finally, $Wc$ is an important component of the $W$ + 1, 2 jet samples, a dataset used in the search of signals such as single top, the Higgs boson and supersymmetric top. In this context, the measurement of the associated production of $W$ bosons with single charm at the Tevatron is an opportunity to test the SM calculation and set an experimental constraint to the theory.

This Letter documents the first measurement of $Wc$ production in $p\bar{p}$ collisions. We use the correlation between the charge of the $W$ and the charge of the muon from the semileptonic decay of the charm hadron in the $W$ + 1, 2 jet sample. Charge conservation in the process  $gq\rightarrow Wc$ ($q=d,s$) allows as final states only the pairings $W^+\bar{c}$ and $W^-c$. It follows that the charge of the lepton from the semileptonic decay of the $c$ quark, and the charge of the lepton from the $W$ decay, are always of opposite sign. The observed lepton charge correlation in $Wc$ is expected to be less than 100\%, due to dilution from hadronic decays in flight and hadrons misidentified as muons (\textquotedblleft mistags\textquotedblright). We identify the $W$ by its decay to an  isolated electron (muon) carrying large transverse energy, $E_T$ (momentum, $p_T$) with respect to the beamline, plus a neutrino. We refer to these high-$p_T$ electrons or muons as \textquotedblleft primary leptons\textquotedblright (PL). The neutrino escapes the detector causing an imbalance of total transverse energy, referred to as \textquotedblleft missing $E_T$\textquotedblright ($\slashed{E}_T$) \cite{cdfsys}. Quarks hadronize and are observed as jets of charged and neutral particles. We identify the charm jets from their semileptonic decay by looking for a muon within the jet  (``soft-lepton tag" or ``SLT"). We use the difference between events in which the PL and SLT are oppositely charged (\textquotedblleft OS\textquotedblright) and events in which they have the same charge (\textquotedblleft SS\textquotedblright) as an observable and define the ``asymmetry": $A=(N_{\rm OS}-N_{\rm SS})/(N_{\rm OS}+N_{\rm SS})$. We measure $A$ and determine the expected number of OS$-$SS events due to background sources, primarily $W+{\rm light~flavor}$ (LF) hadrons, di-muon events from Drell-Yan (DY), and multi-jet events (referred to as ``non-$W$ QCD"). To a smaller extent, single top, $Z\rightarrow\tau\tau$ and $WW$ events are also sources of charge-correlated background. Because the $Wc$ signal produces mainly OS events, whereas most of the background is equally likely to give OS or SS events, we are able to extract the signal from a background-dominated sample using the excess of OS$-$SS events over the expected ({\em not} including $Wc$) number. 

The CDF II detector is described in detail elsewhere \cite{CDF}. The components relevant to this analysis include the silicon vertex detector, the central outer tracker (COT), the central electromagnetic and hadronic calorimeters, the central muon detectors and the luminosity counters. The data sample, produced in $p\bar{p}$ collisions at $\sqrt{s}=1.96~{\rm TeV}$ during Run$~$II of the Fermilab Tevatron, was collected between March 2002 and April 2007. This analysis is based on an integrated luminosity of $1823\pm109~{\rm pb}^{-1}$. Monte Carlo (MC) simulations of $W+{\rm jets}$ production are performed using {\tt ALPGEN} ({\tt v2.1} \cite{alpgen}), coupled with {\tt CTEQ5L} PDFs \cite{cteq5} and {\tt PYTHIA} ({\tt v6.3} \cite{pythia}) for the shower evolution. Modeling of $b$ and $c$ hadron decay is provided by {\tt EvtGen} \cite{evtgen}. DY and $WW$ production are modeled with {\tt PYTHIA} ({\tt v6.5}). Events are selected with an inclusive lepton trigger requiring an electron (muon) with $E_T>18~{\rm GeV}$ ($p_T>18~{\rm GeV}/c$). Further selection requires that candidate PLs are isolated ($I<0.1$ \cite{iso}) and have $E_T$ ($p_T$) greater than 20 GeV with $|\eta|<1.1$. The event must also have $\slashed{E}_T>25$ GeV. Jets are identified using a cone algorithm \cite{jets} with a cone opening of $\Delta R\equiv\sqrt{(\Delta\eta)^2+(\Delta\phi)^2}=0.4$ and are constrained to originate from the $p\bar{p}$ collision vertex. Muons inside jets are identified by matching the tracks of the jet, as measured in the COT, with track segments (\textquotedblleft stubs\textquotedblright) in the muon detectors. Such a muon with $p_T>3$ GeV/$c$ and within $\Delta R<0.6$ of a jet axis is an SLT. The probability of misidentifying a hadron as an SLT muon, denoted as the \textquotedblleft mistag probability\textquotedblright, is measured using jets from data \cite{sltprd}. An SLT mistag probability look-up table is constructed as a function of the track $p_T$, $\eta$ and $\phi$. The number of tags predicted with the look-up table agrees to within 10\% of the observation in a variety of samples, including QCD multijets and $Z+{\rm jet}$ events. To reduce background from di-muon resonances and double-semileptonic $B$ hadron decays we remove events in which the muon PL and SLT are oppositely charged and have an invariant mass consistent with a $Z$, $\Upsilon$ or, irrespectively of the PL flavor, less than 5 ${\rm GeV}/c^2$. Remaining cuts are as in \cite{sltprd}. The jet energies are corrected to account for variations of the detector response in $\eta$ and time, as well as multiple $p\bar{p}$ interactions, but are not corrected back to the parton energy. Finally, the $W+{\rm jet}$ dataset is partitioned according to the number of jets with $E_T>10$ GeV and $|\eta|<2.0$  in the event (corresponding to the charm's $|\eta_c|\lesssim 1.5$, from the geometric acceptance of muons, and $p_{Tc}\gtrsim20~{\rm GeV}/c$). We refer to events in which the PL is an electron (muon) as the ``electron sample" (``muon sample"). In the electron (muon) sample we select 636 (425) OS and 450 (313) SS events with at least one jet tagged by the SLT algorithm. In rare cases where there is more than one tag in the same jet, or more than one tagged jet in the same event, we use the SLT candidate that has the best match between the COT track and muon stubs. 

The composition of the $W+1,~2~{\rm jet}$ sample includes primarily $W+{\rm LF}$, $Wb\bar{b}$, $Wc\bar{c}$, $Wc$, non-$W$ QCD, and DY events. $Wb\bar{b}$ and $Wc\bar{c}$ backgrounds reduce the statistical precision of the measurement but do not contribute to the measured asymmetry, since the presence of both heavy quark and antiquark allows with equal probability either charge combination of the leptons from the $W$ and from the (anti)quark decays. Table~\ref{tab:summary2} shows the data and expected background contributions. The dominant mechanism by which DY events contribute to the background is when one of the produced muons emits a high-energy photon, and the muon-photon pair is identified as a jet, while the muon passes the SLT-tagging requirements. In order to reduce this background we reject events based on the jet's electromagnetic energy fraction and number of tracks. Events with an OS primary muon and SLT that have a jet with less than 2 tracks in a cone of opening $\Delta R=0.6$ around the jet axis and an electromagnetic fraction higher than 0.8 are rejected as potential radiative DY events. After the rejection we estimate the number of remaining DY events in the sample by measuring the number of OS--SS events in the data in a window of invariant mass containing the $Z$ peak, $n^{\rm data}_Z$, and using MC simulation to extrapolate $n^{\rm data}_Z$ in the $Wc$ signal region. The uncertainty shown is due to the size of the data and, to a lesser extent, MC samples.

Events without $W$ bosons that enter the signal sample are typically QCD jet events where one jet has faked a high-$p_T$ lepton, mismeasured energies produce apparent $\slashed{E}_T$ and an additional jet contains an SLT muon. A fraction of these events are from $b\overline{b}$ and $c\overline{c}$, where the PL results from a semileptonic decay on one of the fragmenting heavy quark and the SLT from a semileptonic decay on the other, with a potentially large charge asymmetry. We estimate the number of non-$W$ QCD events in the sample by extrapolating the number of SLT-tagged events with an isolated lepton ($I<0.1$) and $\slashed{E}_T<15~{\rm GeV}$ into the signal region of $\slashed{E}_T>25~{\rm GeV}$. The difference seen in Table \ref{tab:summary2} between the electron and muon channels reflects the higher likelihood for a jet to be misidentified as an electron than as a muon. The extrapolation uncertainty is determined by comparing with data the prediction for the number of events in the region of $0.1<I<0.2$ and $\slashed{E}_T>25~{\rm GeV}$, just outside the signal region. We assign a $\pm20$\% systematic uncertainty based on the difference between the two. We model the charge correlation of non-$W$ events from the data using events with $\slashed{E}_T>25$ GeV but with a non-isolated PL, and the dominant uncertainty is from the limited data sample size.

$W+{\rm LF}$ events enter the data sample when one of the LF tracks in the jet is misidentified as a muon. Since the same process that describes $Wc$ production may be expected to describe $Wu$, we expect a small anti-correlation between the charge of the $W$ and the charge of the tracks in the jets recoiling against the $W$. We estimate the total $W+{\rm LF}$ background using the mistag probability parametrization with the number of events before SLT tagging, correcting for a fraction due to non-W QCD. The uncertainties include contributions from the mistag prediction and from the correction for non-$W$ QCD. We determine the expected asymmetry for these events using {\tt ALPGEN} MC events; the asymmetry is lower in the muon sample as a result of the removal of events consistent with di-muon resonances, which removes only OS events. Remaining small backgrounds with expected asymmetries are from single top ($t$-channel) production, $WW$, and $Z\rightarrow\tau\tau$. These are estimated using theoretical cross sections and MC simulations.  Finally, for $Wb\bar{b}$ and $Wc\bar{c}$ we measure a charge asymmetry from MC events, compatible with zero as expected.

\begin{table}[th]
\caption{{\small Summary of data and charge-asymmetric background in the SLT-tagged $W+1,~2$ jet sample.}}
\begin{center}
\begin{ruledtabular}
\begin{tabular}{lccc}
Source                & Events    &     Asymmetry            & OS--SS events \\
\hline
Drell-Yan               &               &                 & $41.6\pm5.4$ \\
Non-$W$ ($e$ PL)              & $200.0\pm41.5$    & $0.179\pm0.046$ & $35.8\pm11.8$   \\
Non-$W$ ($\mu$ PL)           & $30.2\pm6.5$  & $0.252\pm0.066$ &  $7.6\pm2.6$ \\
$W+{\rm LF}$ ($e$ PL)     & $695.5\pm75.7$ & $0.057\pm0.002$ &  $39.6\pm4.9$ \\
$W+{\rm LF}$ ($\mu$ PL)     & $491.4\pm53.3$ & $0.019\pm0.002$ &  $9.3\pm2.0$ \\
Single top                  &         & & $7.6\pm1.1$  \\
$Z\rightarrow\tau\tau$, $WW$       &        &        &   $7.2\pm1.2$  \\
\hline
Total Background              &               &             & $148.7\pm15.4$  \\
Data                        &    1824        &         & 298  \\
\end{tabular}
\end{ruledtabular}
\end{center}
\label{tab:summary2}
\end{table}

\begin{figure}[th]
\begin{center}
\begin{tabular}{c}
\includegraphics[width=8.5cm]{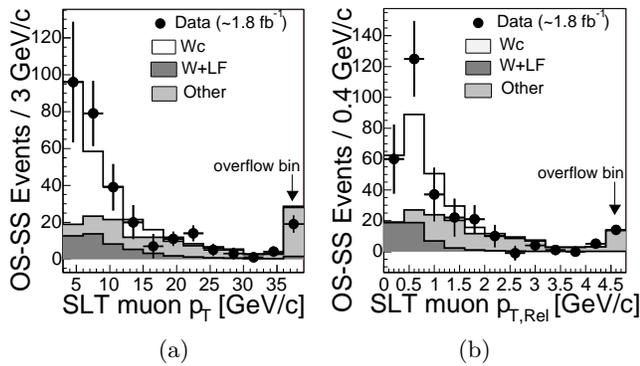} \\
(a) $~~~~~~~~~~~~~~~~~~~~~~~~~~~~~~~$ (b)
\end{tabular}
\end{center}
\caption{{\small SS-subtracted distributions of the SLT muon $p_T$ relative to (a): the beam axis and (b): the SLT-jet axis. The $Wc$ contribution is normalized to this measurement.}}
\label{fig:kindist}
\end{figure}

After background subtraction we measure $149.3\pm42.1({\rm stat})\pm15.4({\rm syst})$ $Wc$ events in the selected SS-subtracted sample. Figure~\ref{fig:kindist} shows two kinematic distributions of the SLT muons in the tagged events. Each of the distributions is SS subtracted and the stacked histograms show the contribution from both the expected $Wc$ signal and the estimated backgrounds. The higher-$p_T$ muons in both plots arise primarily from DY events misidentified as charm decay. The $Wc$ production cross section is obtained using the cross section formula:
\begin{equation}\label{eq:Xsec}
\sigma_{Wc}\times {\rm BR}(W\rightarrow \ell\nu) = \frac{N_{\rm tot}^{OS-SS}-N^{OS-SS}_{\rm bkg}}{Acc
\cdot \int L~{\rm dt}},
\end{equation}

where $\sigma_{Wc}\equiv\sigma_{W^+\bar{c}}+\sigma_{W^-c}$ with $p_{Tc}>20~{\rm GeV}/c$ and $|\eta_c|<1.5$, and ${\rm BR}(W\rightarrow \ell\nu)=0.108$. In the denominator, $Acc$ is the SS-subtracted acceptance times efficiency derived from a MC simulation of $Wc$ events, and $\int L{\rm dt}$ is the integrated luminosity of the data. We find $Acc=(0.833\pm0.010)\times10^{-2}$; the value implicitly includes the expected charge asymmetry of the $Wc$ sample, which is $(70.6\pm0.5)$\%, as well as the semileptonic BR of charm hadrons to muon. The uncertainty is due to the size of the MC sample.
Using Equation~\ref{eq:Xsec}, we find: $\sigma_{Wc}\times {\rm BR}\left(W\rightarrow \ell\nu\right)=9.8\pm2.8({\rm stat})^{+1.4}_{-1.6}({\rm syst})\pm0.6({\rm lum})~{\rm pb}$.

\begin{table}[ht]
\caption{{\small Summary of systematic uncertainties.}}
\begin{center}
\begin{ruledtabular}
\begin{tabular}{lc}
Source                  &   $\Delta\sigma_{Wc}$ (\%)   \\
\hline
Lepton ID    &  $\pm1.4$  \\
SLT tagging efficiency      &  $\pm5.1$  \\
Jet energy calibration       & +4.0/--2.7 \\
PDFs              & $\pm3.6$  \\
$c$-quark hadronization    & $\pm4.6$    \\
ISR, FSR            &  +4.2/--9.5 \\
Factorization scale	  &   $\pm1.3$ \\
Background	     &      $\pm10.3$\\
Luminosity              & $\pm6.0$ \\
\hline
Total (excluding Luminosity)         & +14.2/--16.3\\
\end{tabular}
\end{ruledtabular}
\end{center}
\label{tab:summary3}
\end{table}

Systematic uncertainties are shown in Table~\ref{tab:summary3}. MC modeling of the efficiency for identifying the PL (``lepton ID") is measured using $Z$-boson data and MC samples. The uncertainty on the efficiency for identifying the SLT includes a contribution from MC modeling of particle tracking efficiency in dense jet environments, and a contribution from the parametrization of the SLT efficiency as a function of the $p_T$ of the muon \cite{sltprd}. The PDF uncertainty is derived by using CTEQ and MRST sets, as prescribed in \cite{topmass}. We compare charm jets in {\tt PYTHIA} and {\tt HERWIG} \cite{herwig1,herwig2} to evaluate the uncertainty due to different hadronization models. To measure the effects of initial (ISR) and final state radiation (FSR) we use inclusive $Wc$ production in {\tt ALPGEN} and compare the nominal acceptance with enhanced or reduced radiation as in \cite{topmass}. The uncertainty on the factorization and renormalization scales is estimated by varying them in {\tt ALPGEN} between the transverse mass of the $W$ and that of the charm. Finally, uncertainties on the background estimations are included in the cross section as systematic uncertainties. 

In conclusion, we present the first measurement of the $Wc$ production cross section using $\sim1.8~{\rm fb}^{-1}$ of $p\bar{p}$ collisions at $\sqrt{s}=1.96~{\rm TeV}$. We measure $\sigma_{Wc}\left(p_{Tc}>20~{\rm GeV}/c,~|\eta_c|<1.5\right)\times {\rm BR}\left(W\rightarrow\ell\nu\right) = 9.8\pm3.2~{\rm pb}$, in agreement with the NLO calculation over the same phase space region of: $11.0^{+1.4}_{-3.0}~{\rm pb}$ \cite{campbell}. The result provides an experimental validation of the theoretical $Wc$ cross section for use in single top and Higgs search analyses. Within the quoted sensitivity, we see no evidence of the presence of an exotic particle with high semileptonic BR. Finally, the result provides the first experimental constraints to the theory uncertainties; an extrapolation indicates that a precision of $\sim15$\% can be achieved with $\sim$~6--7 fb$^{-1}$ of data, expected by the end of Run II.

We thank the Fermilab staff and the technical staffs of the participating institutions for their vital contributions. This work was supported by the U.S. Department of Energy and National Science Foundation; the Italian Istituto Nazionale di Fisica Nucleare; the Ministry of Education, Culture, Sports, Science and Technology of Japan; the Natural Sciences and Engineering Research Council of Canada; the National Science Council of the Republic of China; the Swiss National Science Foundation; the A.P. Sloan Foundation; the Bundesministerium f\"ur Bildung und Forschung, Germany; the Korean Science and Engineering Foundation and the Korean Research Foundation; the Science and Technology Facilities Council and the Royal Society, UK; the Institut National de Physique Nucleaire et Physique des Particules/CNRS; the Russian Foundation for Basic Research; the Comisi\'on Interministerial de Ciencia y Tecnolog\'{\i}a, Spain; the European Community's Human Potential Programme; the Slovak R\&D Agency; and the Academy of Finland.

\end{document}

%% file: July_2007_Authors_Visitors.tex
\affiliation{Institute of Physics, Academia Sinica, Taipei, Taiwan 11529, Republic of China} 
\affiliation{Argonne National Laboratory, Argonne, Illinois 60439} 
\affiliation{Institut de Fisica d'Altes Energies, Universitat Autonoma de Barcelona, E-08193, Bellaterra (Barcelona), Spain} 
\affiliation{Baylor University, Waco, Texas  76798} 
\affiliation{Istituto Nazionale di Fisica Nucleare, University of Bologna, I-40127 Bologna, Italy} 
\affiliation{Brandeis University, Waltham, Massachusetts 02254} 
\affiliation{University of California, Davis, Davis, California  95616} 
\affiliation{University of California, Los Angeles, Los Angeles, California  90024} 
\affiliation{University of California, San Diego, La Jolla, California  92093} 
\affiliation{University of California, Santa Barbara, Santa Barbara, California 93106} 
\affiliation{Instituto de Fisica de Cantabria, CSIC-University of Cantabria, 39005 Santander, Spain} 
\affiliation{Carnegie Mellon University, Pittsburgh, PA  15213} 
\affiliation{Enrico Fermi Institute, University of Chicago, Chicago, Illinois 60637} 
\affiliation{Comenius University, 842 48 Bratislava, Slovakia; Institute of Experimental Physics, 040 01 Kosice, Slovakia} 
\affiliation{Joint Institute for Nuclear Research, RU-141980 Dubna, Russia} 
\affiliation{Duke University, Durham, North Carolina  27708} 
\affiliation{Fermi National Accelerator Laboratory, Batavia, Illinois 60510} 
\affiliation{University of Florida, Gainesville, Florida  32611} 
\affiliation{Laboratori Nazionali di Frascati, Istituto Nazionale di Fisica Nucleare, I-00044 Frascati, Italy} 
\affiliation{University of Geneva, CH-1211 Geneva 4, Switzerland} 
\affiliation{Glasgow University, Glasgow G12 8QQ, United Kingdom} 
\affiliation{Harvard University, Cambridge, Massachusetts 02138} 
\affiliation{Division of High Energy Physics, Department of Physics, University of Helsinki and Helsinki Institute of Physics, FIN-00014, Helsinki, Finland} 
\affiliation{University of Illinois, Urbana, Illinois 61801} 
\affiliation{The Johns Hopkins University, Baltimore, Maryland 21218} 
\affiliation{Institut f\"{u}r Experimentelle Kernphysik, Universit\"{a}t Karlsruhe, 76128 Karlsruhe, Germany} 
\affiliation{Center for High Energy Physics: Kyungpook National University, Daegu 702-701, Korea; Seoul National University, Seoul 151-742, Korea; Sungkyunkwan University, Suwon 440-746, Korea; Korea Institute of Science and Technology Information, Daejeon, 305-806, Korea; Chonnam National University, Gwangju, 500-757, Korea} 
\affiliation{Ernest Orlando Lawrence Berkeley National Laboratory, Berkeley, California 94720} 
\affiliation{University of Liverpool, Liverpool L69 7ZE, United Kingdom} 
\affiliation{University College London, London WC1E 6BT, United Kingdom} 
\affiliation{Centro de Investigaciones Energeticas Medioambientales y Tecnologicas, E-28040 Madrid, Spain} 
\affiliation{Massachusetts Institute of Technology, Cambridge, Massachusetts  02139} 
\affiliation{Institute of Particle Physics: McGill University, Montr\'{e}al, Canada H3A~2T8; and University of Toronto, Toronto, Canada M5S~1A7} 
\affiliation{University of Michigan, Ann Arbor, Michigan 48109} 
\affiliation{Michigan State University, East Lansing, Michigan  48824} 
\affiliation{University of New Mexico, Albuquerque, New Mexico 87131} 
\affiliation{Northwestern University, Evanston, Illinois  60208} 
\affiliation{The Ohio State University, Columbus, Ohio  43210} 
\affiliation{Okayama University, Okayama 700-8530, Japan} 
\affiliation{Osaka City University, Osaka 588, Japan} 
\affiliation{University of Oxford, Oxford OX1 3RH, United Kingdom} 
\affiliation{University of Padova, Istituto Nazionale di Fisica Nucleare, Sezione di Padova-Trento, I-35131 Padova, Italy} 
\affiliation{LPNHE, Universite Pierre et Marie Curie/IN2P3-CNRS, UMR7585, Paris, F-75252 France} 
\affiliation{University of Pennsylvania, Philadelphia, Pennsylvania 19104} 
\affiliation{Istituto Nazionale di Fisica Nucleare Pisa, Universities of Pisa, Siena and Scuola Normale Superiore, I-56127 Pisa, Italy} 
\affiliation{University of Pittsburgh, Pittsburgh, Pennsylvania 15260} 
\affiliation{Purdue University, West Lafayette, Indiana 47907} 
\affiliation{University of Rochester, Rochester, New York 14627} 
\affiliation{The Rockefeller University, New York, New York 10021} 
\affiliation{Istituto Nazionale di Fisica Nucleare, Sezione di Roma 1, University of Rome ``La Sapienza," I-00185 Roma, Italy} 
\affiliation{Rutgers University, Piscataway, New Jersey 08855} 
\affiliation{Texas A\&M University, College Station, Texas 77843} 
\affiliation{Istituto Nazionale di Fisica Nucleare, University of Trieste/\ Udine, Italy} 
\affiliation{University of Tsukuba, Tsukuba, Ibaraki 305, Japan} 
\affiliation{Tufts University, Medford, Massachusetts 02155} 
\affiliation{Waseda University, Tokyo 169, Japan} 
\affiliation{Wayne State University, Detroit, Michigan  48201} 
\affiliation{University of Wisconsin, Madison, Wisconsin 53706} 
\affiliation{Yale University, New Haven, Connecticut 06520} 
\author{T.~Aaltonen}
\affiliation{Division of High Energy Physics, Department of Physics, University of Helsinki and Helsinki Institute of Physics, FIN-00014, Helsinki, Finland}
\author{J.~Adelman}
\affiliation{Enrico Fermi Institute, University of Chicago, Chicago, Illinois 60637}
\author{T.~Akimoto}
\affiliation{University of Tsukuba, Tsukuba, Ibaraki 305, Japan}
\author{M.G.~Albrow}
\affiliation{Fermi National Accelerator Laboratory, Batavia, Illinois 60510}
\author{B.~\'{A}lvarez~Gonz\'{a}lez}
\affiliation{Instituto de Fisica de Cantabria, CSIC-University of Cantabria, 39005 Santander, Spain}
\author{S.~Amerio}
\affiliation{University of Padova, Istituto Nazionale di Fisica Nucleare, Sezione di Padova-Trento, I-35131 Padova, Italy}
\author{D.~Amidei}
\affiliation{University of Michigan, Ann Arbor, Michigan 48109}
\author{A.~Anastassov}
\affiliation{Rutgers University, Piscataway, New Jersey 08855}
\author{A.~Annovi}
\affiliation{Laboratori Nazionali di Frascati, Istituto Nazionale di Fisica Nucleare, I-00044 Frascati, Italy}
\author{J.~Antos}
\affiliation{Comenius University, 842 48 Bratislava, Slovakia; Institute of Experimental Physics, 040 01 Kosice, Slovakia}
% ...
\collaboration{CDF Collaboration\footnote{With visitors from $^a$University of Athens, 15784 Athens, Greece, 
$^b$Chinese Academy of Sciences, Beijing 100864, China, 
$^c$University of Bristol, Bristol BS8 1TL, United Kingdom, 
$^d$University Libre de Bruxelles, B-1050 Brussels, Belgium, 
$^e$University of California Irvine, Irvine, CA  92697, 
$^f$University of California Santa Cruz, Santa Cruz, CA  95064, 
$^g$Cornell University, Ithaca, NY  14853, 
$^h$University of Cyprus, Nicosia CY-1678, Cyprus, 
$^i$University College Dublin, Dublin 4, Ireland, 
$^j$University of Edinburgh, Edinburgh EH9 3JZ, United Kingdom, 
$^k$University of Heidelberg, D-69120 Heidelberg, Germany, 
$^l$Universidad Iberoamericana, Mexico D.F., Mexico, 
$^m$University of Manchester, Manchester M13 9PL, England, 
$^n$Nagasaki Institute of Applied Science, Nagasaki, Japan, 
$^o$University de Oviedo, E-33007 Oviedo, Spain, 
$^p$Queen Mary, University of London, London, E1 4NS, England, 
$^q$Texas Tech University, Lubbock, TX  79409, 
$^r$IFIC(CSIC-Universitat de Valencia), 46071 Valencia, Spain, 
$^s$European Organization for Nuclear Research CERN, CH-1211 Geneve 23, Switzerland.
}}
\noaffiliation

%% file: wcprl.bbl
\begin{thebibliography}{99.}
\bibitem{cdfsys} We use a ($z,\phi, \theta$) coordinate system where the $z$-axis is in the direction of the proton beam, and $\phi$ and $\theta$ are the azimuthal and polar angles respectively.  The pseudo-rapidity is $\eta\equiv-\ln(\tan\frac{\theta}{2})$. Transverse energy and momentum are $E_T\equiv E\sin\theta$ and $p_T\equiv p\sin\theta$ respectively, where $E$ and $p$ are energy and momentum. The missing transverse energy is $\slashed{E}_T \equiv |-\sum_i E_T^i\hat{n}_i|$, where $E_T^i$ is the magnitude of the transverse energy contained in each calorimeter tower $i$ in the region $|\eta|<3.6$, and $\hat{n}_i$ is the direction unit vector of the tower in the plane transverse to the beam direction.
\bibitem{iso} The isolation ($I$) is defined as the calorimeter transverse energy in a cone of opening $\Delta R\equiv \sqrt{(\Delta\eta)^2+(\Delta\phi)^2}=0.4$ around the lepton (not including the lepton energy itself) divided by the lepton $E_T$ or $p_T$.
\end{thebibliography}
